\pdfoutput=1
\documentclass{article}
\usepackage[preprint]{spconfa4,amsmath,graphicx}

\usepackage{multirow}
\usepackage{pifont}
\usepackage{booktabs}
\usepackage{amssymb}
\usepackage{amsmath}
\usepackage{tikz}
\usepackage{enumitem}
\usepackage{xspace}
\usepackage{pgfplots}\usepackage[acronym, toc,nonumberlist]{glossaries}
\usepackage{balance}
\usepackage{xifthen}
\usepackage{hyperref}
\usepackage{url}
\usepackage[capitalize]{cleveref}

\usepackage{siunitx}
\robustify\bfseries  
\usepackage{amsmath,graphicx}
\usetikzlibrary{calc}
\usetikzlibrary{tikzmark}
\usetikzlibrary{arrows, positioning}
\usetikzlibrary{fit}
\usepgfplotslibrary[groupplots]
\pgfplotsset{compat=newest}
\usepackage[shortcuts]{extdash}

\usepackage{csquotes}
\usepackage{subcaption}

\usepackage{url}

\newcolumntype{H}{>{\setbox0=\hbox\bgroup}c<{\egroup}@{}}   

\newcommand*{\thl}{\fontseries{b}\selectfont}
\robustify\thl
\robustify\fontseries

\makeatletter
\renewcommand\section{\@startsection {section}{1}{\z@}%
                                  {-3.5ex \@plus -1ex \@minus -.2ex}%
                                  {2.3ex \@plus.2ex}%
                                  {\normalfont\Large\bfseries}}
\renewcommand\subsection{\@startsection{subsection}{2}{\z@}%
                                     {-3.25ex\@plus -1ex \@minus -.2ex}%
                                     {1ex \@plus .2ex}%
                                     {\normalfont\large\bfseries}}
\renewcommand\subsubsection{\@startsection{subsubsection}{3}{\z@}%
                                     {-2.25ex\@plus -1ex \@minus -.2ex}%
                                     {0.25ex \@plus .2ex}%
                                     {\normalfont\normalsize\bfseries}}
\renewcommand\paragraph{\@startsection{paragraph}{4}{\z@}%
                                    {3.25ex \@plus1ex \@minus.2ex}%
                                    {-1em}%
                                    {\normalfont\normalsize\bfseries}}
\renewcommand\subparagraph{\@startsection{subparagraph}{5}{\parindent}%
                                      {3.25ex \@plus1ex \@minus .2ex}%
                                      {-1em}%
                                      {\normalfont\normalsize\bfseries}}
\makeatother
\usepackage{tikz}
\usepackage{pgfplots}
\pgfplotsset{compat=1.16}
\usepgfplotslibrary{groupplots}
\usetikzlibrary{positioning,arrows,matrix,fit,calc,patterns,chains,scopes,shapes.multipart,decorations,decorations.markings,backgrounds}

\definecolor{palette-1}{HTML}{001C7F}    
\definecolor{palette-2}{HTML}{B1400D}    
\definecolor{palette-3}{HTML}{12711C}    
\definecolor{palette-4}{HTML}{8C0800}    
\definecolor{palette-5}{HTML}{591E71}    
\definecolor{palette-6}{HTML}{592F0D}    
\definecolor{palette-7}{HTML}{A23582}    
\definecolor{palette-8}{HTML}{3C3C3C}    
\definecolor{palette-9}{HTML}{B8850A}    
\definecolor{palette-10}{HTML}{006374}      

\tikzset{
    line/.style={draw,black,thick,rounded corners=1mm,line cap=round},
    noshortarrow/.style={line,->},
    arrow/.style={noshortarrow,shorten >=.3mm},
    doublearrow/.style={arrow,<->, shorten <=.3mm},
    box/.style={draw,black,minimum height=3em,text height=1.5ex,text depth=0.25ex,rounded corners=2,fill=white},
    nopadding/.style={minimum height=0,inner sep=1mm},
    pbox/.style={box,fill=black!10},
    backgroundbox/.style={inner xsep=3mm, inner ysep=1mm, draw, dashed, rounded corners,fill=orange!10},
    branch/.style={inner sep=0.3mm,circle,fill=black},
    operator/.style={draw,circle,black,rounded corners,inner sep=0,fill=white},
    pattern1/.style={pattern=north west lines,pattern color=palette-1},
    pattern2/.style={pattern=north east lines,pattern color=palette-2},
    pattern3/.style={pattern=crosshatch,pattern color=palette-3},
    utt/.style={box,minimum height=1mm,inner sep=0,rounded corners=0.5,text=white,font=\footnotesize\sffamily},
    utt1/.style={utt,fill=palette-1, minimum width=1.5cm},
    utt2/.style={utt,fill=palette-2, minimum width=1cm},
    utt3/.style={utt,fill=palette-3, minimum width=1cm},
    utt4/.style={utt,fill=palette-4, minimum width=1cm},
    overlap/.style={fill=yellow,opacity=0.5},
    uttbox/.style={fill=black!15,draw=none,inner xsep=0,rounded corners=3},
}

\newcommand\defm[2]{\expandafter\newcommand{#1}{\ensuremath{#2}}} 

\defm\conv{*}

\defm\mix{y}
\defm\image{x}
\defm\source{s}
\defm\rir{h}
\defm\noise{\nu}
\defm\offset{\tau}
\defm\timeframe{t}
\defm\utt{n}
\defm\nutt{N}
\defm\scale{\alpha}
\defm\nspk{K}
\defm\spk{k}
\defm\activity{\pi}
\defm\estactivity{\hat{\activity}}
\defm\prob{p}

\let\oldcite\cite
\renewcommand{\cite}[1]{%
\ifthenelse{\isempty{#1}}%
{\inred{[cite!]}}%
{\oldcite{#1}}%
}



\newacronym{VAD}{VAD}{Voice Activity Detection}
\newacronym{SASDR}{SA-SDR}{Source-Aggregated Signal-to-Distortion Ratio}
\newacronym{WER}{WER}{Word Error Rate}
\newacronym{SDR}{SDR}{Signal-to-Distortion Ratio}
\newacronym{TDNN-F}{TDNN-F}{Factorized Time-Delayed Neural Network}
\newacronym{MMS-MSG}{MMS-MSG}{Multi-Purpose Multi-Speaker Mixture Signal Generator}
\newacronym{ORC-WER}{ORC-WER}{Optimal Reference Combination Word Error Rate}
\newacronym{ASR}{ASR}{Automatic Speech Recognition}
\newacronym{RIR}{RIR}{Room Impulse Response}
\newacronym{BLSTM}{BLSTM}{Bidirectional Long Short-Term Memory}
\newacronym{CSS}{CSS}{Continuous Speech Separation}

\copyrightnotice{978-1-6654-6867-1/22/\$31.00~\copyright2022 IEEE}
                   
\title{MMS-MSG: A Multi-purpose Multi-Speaker Mixture Signal Generator}
\name{Tobias Cord-Landwehr, Thilo von Neumann, Christoph Boeddeker, Reinhold Haeb-Umbach}%
\address{Paderborn University, Department of Communications Engineering, Paderborn, Germany}
\begin{document}
\ninept
\maketitle
\begin{abstract}
    The scope of speech enhancement has changed from a monolithic view of single, independent tasks, to a joint processing of complex conversational speech recordings.
    Training and evaluation of these single tasks requires synthetic data with access to intermediate signals that is as close as possible to the evaluation scenario.
    As such data often is not available, many works instead use specialized databases for the training of each system component, e.g WSJ0-mix for source separation.
    We present a Multi-purpose Multi-Speaker Mixture Signal Generator (MMS-MSG) for generating a variety of speech mixture signals based on any speech corpus, ranging from classical anechoic mixtures (e.g., WSJ0-mix) over reverberant mixtures (e.g., SMS-WSJ) to meeting-style data.
    Its highly modular and flexible structure allows for the simulation of diverse environments and dynamic mixing, while simultaneously enabling an easy extension and modification to generate new scenarios and mixture types. These meetings can be used for prototyping, evaluation, or training purposes. 
    We provide example evaluation data and baseline results for meetings based on the WSJ corpus. Further, we demonstrate the usefulness for realistic scenarios by using MMS-MSG to provide training data for the LibriCSS database.

  
  
  
  
  

\end{abstract}


\section{Introduction}

Multi-talker conversational speech recognition is concerned with the transcription of audio recordings of formal meetings or informal get-togethers in machine-readable form using distant microphones. 
The exhaustive task is to obtain a transcription for each present speaker.
Often, there is the need or desire to enrich the output with a diarization component that delivers information about who spoke when.

To solve the transcription task, several speech processing components have to be employed:
i) speech enhancement to reduce the impact of reverberation, environmental noise or remaining residuals from an interference on the transcription performance of the system,
ii) a source separation module that decomposes overlapped speech into the signals of the individual speakers. This latter module is necessary, because it has been observed that in a significant amount of time, in the order of \SIrange{5}{20}{\percent}, more than one participant is speaking.
Furthermore, there often is iii) a diarization component and iv) an ASR back-end.
Traditionally, all these tasks have been considered separately, and only in recent years there is a trend to solve them jointly, either by employing monolithic neural network architectures \cite{Sklyar2021_MultiturnRNNTStreaming, kanda2022e2easr} or by relying on separate processing components that are optimized separately, but nevertheless used jointly \cite{21_raj_jsalt, 20_Chen_libricss}.

As those speech processing components mainly are deep learning systems,
important prerequisites of system development are appropriate training and evaluation databases. For each of the above components there exist such databases, often designed in the context of community challenges, such as i) REVERB \cite{13_kinoshita_reverb}, DNS \cite{21_chandan_dns},  CHiME-3/4: ii) WSJ0-2/3mix \cite{16_Hershey_dc_wsjmix}, LibriMix \cite{20_consentino_librimix}, SMS-WSJ \cite{Drude_19_smswsj}; iii) DiHARD \cite{20_ryant_dihard}, VoxCeleb \cite{19_nagrani_voxceleb}, VoxConverse \cite{20_chung_voxconverse}, CallHome, and iv) LibriSpeech \cite{Panayotov2015_LibrispeechASRCorpus} and many other.

There also exist databases of real recordings of meeting data, such as AMI \cite{05_mccowan_ami} and CHiME-5/6 \cite{18_barker_chime5}. Furthermore, the LibriCSS \cite{20_Chen_libricss} database contains re-recordings of loudspeaker playback of LibriSpeech sentences mixed to reflect a typical meeting scenario. While being an excellent testbed for system evaluation, real meeting recordings are often unsuitable for system development and training. That is because the training of several system components requires clean target signals for supervised learning. Also, clean references signals are instrumental to be able to assess the performance of individual system components, e.g., the source separation component, and thus pinpoint performance bottlenecks in the overall processing chain. Looking at the final Word Error Rate may not reveal such important diagnostic information. 

This contribution is meant to fill this gap: we introduce \gls{MMS-MSG}, an open-source software for the generation of databases for the training and prototyping of meeting recognition systems. 
While \cite{22_yamashita_dia_data} recently proposed a way to simulate meeting data for diarization systems in particular, this software is designed to provide a highly modular framework that supports the creation of all commonly needed mixture scenarios.
It gives the system developer full flexibility to develop and test meeting recognition systems under various conditions. Key properties are:
\begin{itemize}[nosep,leftmargin=1em,labelwidth=*,align=left, noitemsep]
    \item arbitrary input databases from where to draw the speech samples (e.g., WSJ, LibriSpeech),
    \item generation of single- or multi-channel signals,
    \item generation of anechoic or reverberant signals,
    \item generation of meeting-like speech with a user-defined number of speakers, percentage of speech from individual speakers, and degree of overlapping speech,
    \item native support of dynamic mixing, on-the-fly data generation and quick prototyping,
    \item access to ground truth (e.g., transcription, speaker-id, clean source signals) and intermediate signals (e.g., reverberated source signals), and 
    \item generation of mixtures for common databases to extend the training data with dynamic mixing.
\end{itemize}
We note, though, that, as long as the source signals are taken from non-meeting databases such as WSJ, the generated data can not mimic meetings in the (language-model) sense that there is no real discussion. However, since the acoustic properties, not the content are crucial for most systems,  \gls{MMS-MSG} serves to create meeting scenarios nevertheless.

In this paper we describe the design rationale and methodology, the key properties of the database, and offer results of a baseline system.
Source code to compile the database or to generate data on the fly is released under \url{https://github.com/fgnt/mms_msg}.

\section{Characteristics of meeting-style data}
\label{sec:meeting_characteristics}
In this work, the term meeting is to be
understood in a broad sense, ranging from professional meetings to informal get-togethers
among friends. Speech recorded in meeting scenarios has unique properties, which are quite different to characteristics of other applications:
\begin{itemize}[nosep,leftmargin=1em,labelwidth=*,align=left, noitemsep]
\item Challenging recording conditions: The speech signal is captured in an enclosure by microphones from a distance and is therefore reverberated and often contains  acoustic environmental noise.
\item  Partly overlapped speech: it has been observed that the  time segments where more than one person is speaking, is on the order of 5\% to 10\% \cite{05_mccowan_ami}, while in informal get-togethers it can even exceed 20\% \cite{18_barker_chime5}. Thus, speech separation is a relevant problem. However, the data is different to what is typically studied in source separation, where speakers are fully overlapping.
\item The interaction dynamics of the scenario: There is a limited number of speakers, and speakers are not active continuously. 
Speakers articulate themselves in an intermittent manner with alternating segments of speech inactivity, single-, and multi-talker
speech. Depending on the type of meeting, one or a few speakers may have a significantly larger share of speaking time compared to others.
\item Changing speaker positions: Over the course of a meeting, the speakers may move or at least turn their heads, leading to a changing acoustic transfer function over time. 
\end{itemize}

\begin{figure}[bt]
    \begin{subfigure}[b]{0.3\columnwidth}
        \centering
        \begin{tikzpicture}
    \node[utt1] at (0, 0) (utt1) {A};
    \node[utt2,anchor=north west] at ($(utt1.south west) + (0,-1mm)$) (utt2) {B};
    \draw[dashed] ([yshift=1mm]utt1.north west) -- ([yshift=-1mm]utt2.south west) node[below] {$\offset_1=\offset_2$};
    \begin{pgfonlayer}{background}
        \node[fit={(utt1) (utt2)},uttbox] (uttbox) {};
        \path[overlap] (uttbox.north-|utt2.west) rectangle (uttbox.south-|utt2.east);
    \end{pgfonlayer}
\end{tikzpicture}
        \caption[]{WSJ0-2mix}
    \end{subfigure}
    \begin{subfigure}[b]{0.3\columnwidth}
        \centering
        \begin{tikzpicture}
    \node[utt1] at (0, -2cm) (utt1) {A};
    \node[utt2,anchor=north west] at ($(utt1.south west) + (0.3cm,-1mm)$) (utt2) {B};
    \draw[dashed] ([yshift=1mm]utt1.north west) -- ([yshift=-1mm]utt2.south-|utt1.west) node[below] {$\offset_1$};
    \draw[dashed] ([yshift=1mm]utt1.north-|utt2.west) -- ([yshift=-1mm]utt2.south-|utt2.west) node[below] {$\offset_2$};
    \begin{pgfonlayer}{background}
        \node[fit={(utt1) (utt2)},uttbox] (uttbox) {};
        \path[overlap] (uttbox.north-|utt2.west) rectangle (uttbox.south-|utt2.east);
    \end{pgfonlayer}
\end{tikzpicture}
        \caption[]{SMS-WSJ}
    \end{subfigure}
    \begin{subfigure}[b]{0.3\columnwidth}
        \centering
        \begin{tikzpicture}
    \node[utt1] at (0, 0) (utt1) {A};
    \node[utt2,anchor=north west] at ($(utt1.south west) + (0.9cm,-1mm)$) (utt2) {B};
    \draw[dashed] ([yshift=1mm]utt1.north west) -- ([yshift=-1mm]utt2.south-|utt1.west) node[below] {$\offset_1$};
    \draw[dashed] ([yshift=1mm]utt1.north-|utt2.west) -- ([yshift=-1mm]utt2.south-|utt2.west) node[below] {$\offset_2$};
    \begin{pgfonlayer}{background}
        \node[fit={(utt1) (utt2)},uttbox] (uttbox) {};
        \path[overlap] (uttbox.north-|utt2.west) rectangle (uttbox.south-|utt1.east);
    \end{pgfonlayer}
\end{tikzpicture}
        \caption[]{Partial overlap \cite{Yoshioka2018multi}}
    \end{subfigure}
    \caption{Illustration of the speech mixture scenarios commonly used for source separation systems.}
    \label{fig:single_mixture}
\end{figure}
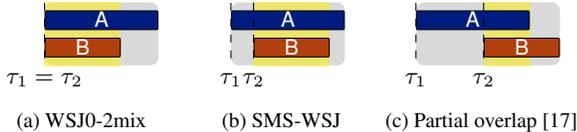

In the next Section we discuss how we addressed these specifics in the design of the mixture signal generator such that both the typical source separation scenarios depicted in \cref{fig:single_mixture} as well as meeting scenarios can be simulated.

\section{Data Simulation}
\label{sec:simulation}


\subsection{Signal Model}
Each mixture signal $\mix(\timeframe)$ that is generated by \gls{MMS-MSG} is modelled as a sum of $\nutt$ utterance signals $\source_\utt(\timeframe)$, uttered by $\nspk$ different speakers, with different start time points $\offset_\utt$, each scaled by a factor $\scale_\utt$ and reverberated with a finite room impulse response $\rir_\utt(\timeframe)$ to model the spatial properties of a scenario:
\begin{align}
    \mix(\timeframe) = \sum_{\utt=1}^\nutt \scale_\utt \source_\utt(\timeframe-\offset_\utt)\conv \rir_\utt(\timeframe) + \noise(\timeframe).
    \label{eq:signal-model}
\end{align}
We make the assumption that the scaling factors and room impulse responses are constant within an utterance, but may change between utterances, e.g., due to speaker movement.
The utterance signals $\source_\utt(\timeframe)$ are zero padded, i.e., any samples that lie outside actual speech activity are $0$.
All distortions to the mixture signal are modelled by an additive noise term $\noise(\timeframe)$ which can consist of different kinds of noise, e.g., white microphone noise or more complex environmental distortions.
Anechoic mixtures can be simulated by omitting the convolution with the \gls{RIR} $\rir_\utt(\timeframe)$. 

\subsection{Simulation Process}

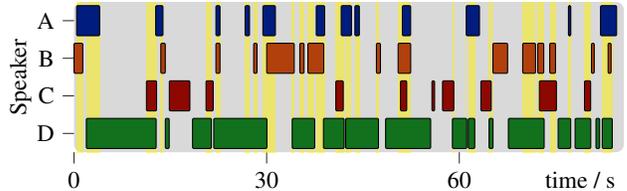
\begin{figure}[bt]
    \centering
\begin{tikzpicture}

\definecolor{darkslategray38}{RGB}{38,38,38}
\definecolor{gray}{RGB}{128,128,128}
\definecolor{green}{RGB}{0,128,0}
\definecolor{lightgray204}{RGB}{204,204,204}
\definecolor{orange}{RGB}{255,165,0}

\begin{axis}[
    scale only axis,
    width=0.85\columnwidth,
    height=2cm,
    axis line style={rounded corners=3,draw=none},
    axis background/.style={uttbox},
    tick align=outside,
    xtick pos=bottom,
    xmin=0, xmax=1370000,
    xtick={0,480000, 960000, 1440000,1920000},
    xticklabels={0,30,60,90,120},
    xminorgrids,
    xtick style={color=darkslategray38},
    y grid style={lightgray204},
    ymin=-0.5, ymax=3.5,
    ytick style={color=darkslategray38},
    scaled x ticks=base 10:0,
    ytick={0,1,2,3,4},
    ytick pos=left,
    yticklabels={D, C, B, A},
    ylabel={Speaker},
    ylabel shift=-4pt,
    xlabel={time / s},
    xlabel style={
        at={(ticklabel cs:1)},
        anchor={south east}
    }
]
\path [overlap]
(axis cs:6560,-0.5)
--(axis cs:6560,3.5)
--(axis cs:21920,3.5)
--(axis cs:21920,-0.5)
--cycle;
\path [overlap]
(axis cs:30720,-0.5)
--(axis cs:30720,3.5)
--(axis cs:63840,3.5)
--(axis cs:63840,-0.5)
--cycle;
\path [overlap]
(axis cs:180480,-0.5)
--(axis cs:180480,3.5)
--(axis cs:205760,3.5)
--(axis cs:205760,-0.5)
--cycle;
\path [overlap]
(axis cs:216320,-0.5)
--(axis cs:216320,3.5)
--(axis cs:221600,3.5)
--(axis cs:221600,-0.5)
--cycle;
\path [overlap]
(axis cs:328960,-0.5)
--(axis cs:328960,3.5)
--(axis cs:342240,3.5)
--(axis cs:342240,-0.5)
--cycle;
\path [overlap]
(axis cs:353600,-0.5)
--(axis cs:353600,3.5)
--(axis cs:363680,3.5)
--(axis cs:363680,-0.5)
--cycle;
\path [overlap]
(axis cs:426240,-0.5)
--(axis cs:426240,3.5)
--(axis cs:437280,3.5)
--(axis cs:437280,-0.5)
--cycle;
\path [overlap]
(axis cs:447680,-0.5)
--(axis cs:447680,3.5)
--(axis cs:456000,3.5)
--(axis cs:456000,-0.5)
--cycle;
\path [overlap]
(axis cs:470560,-0.5)
--(axis cs:470560,3.5)
--(axis cs:501760,3.5)
--(axis cs:501760,-0.5)
--cycle;
\path [overlap]
(axis cs:543680,-0.5)
--(axis cs:543680,3.5)
--(axis cs:548640,3.5)
--(axis cs:548640,-0.5)
--cycle;
\path [overlap]
(axis cs:562560,-0.5)
--(axis cs:562560,3.5)
--(axis cs:573120,3.5)
--(axis cs:573120,-0.5)
--cycle;
\path [overlap]
(axis cs:582720,-0.5)
--(axis cs:582720,3.5)
--(axis cs:599680,3.5)
--(axis cs:599680,-0.5)
--cycle;
\path [overlap]
(axis cs:603520,-0.5)
--(axis cs:603520,3.5)
--(axis cs:624480,3.5)
--(axis cs:624480,-0.5)
--cycle;
\path [overlap]
(axis cs:652320,-0.5)
--(axis cs:652320,3.5)
--(axis cs:673120,3.5)
--(axis cs:673120,-0.5)
--cycle;
\path [overlap]
(axis cs:676800,-0.5)
--(axis cs:676800,3.5)
--(axis cs:691360,3.5)
--(axis cs:691360,-0.5)
--cycle;
\path [overlap]
(axis cs:699840,-0.5)
--(axis cs:699840,3.5)
--(axis cs:710880,3.5)
--(axis cs:710880,-0.5)
--cycle;
\path [overlap]
(axis cs:753440,-0.5)
--(axis cs:753440,3.5)
--(axis cs:758560,3.5)
--(axis cs:758560,-0.5)
--cycle;
\path [overlap]
(axis cs:807520,-0.5)
--(axis cs:807520,3.5)
--(axis cs:838880,3.5)
--(axis cs:838880,-0.5)
--cycle;
\path [overlap]
(axis cs:942880,-0.5)
--(axis cs:942880,3.5)
--(axis cs:946240,3.5)
--(axis cs:946240,-0.5)
--cycle;
\path [overlap]
(axis cs:977280,-0.5)
--(axis cs:977280,3.5)
--(axis cs:978560,3.5)
--(axis cs:978560,-0.5)
--cycle;
\path [overlap]
(axis cs:982400,-0.5)
--(axis cs:982400,3.5)
--(axis cs:998400,3.5)
--(axis cs:998400,-0.5)
--cycle;
\path [overlap]
(axis cs:1034080,-0.5)
--(axis cs:1034080,3.5)
--(axis cs:1039520,3.5)
--(axis cs:1039520,-0.5)
--cycle;
\path [overlap]
(axis cs:1043680,-0.5)
--(axis cs:1043680,3.5)
--(axis cs:1044000,3.5)
--(axis cs:1044000,-0.5)
--cycle;
\path [overlap]
(axis cs:1117920,-0.5)
--(axis cs:1117920,3.5)
--(axis cs:1149920,3.5)
--(axis cs:1149920,-0.5)
--cycle;
\path [overlap]
(axis cs:1155360,-0.5)
--(axis cs:1155360,3.5)
--(axis cs:1171360,3.5)
--(axis cs:1171360,-0.5)
--cycle;
\path [overlap]
(axis cs:1185280,-0.5)
--(axis cs:1185280,3.5)
--(axis cs:1199680,3.5)
--(axis cs:1199680,-0.5)
--cycle;
\path [overlap]
(axis cs:1231840,-0.5)
--(axis cs:1231840,3.5)
--(axis cs:1237600,3.5)
--(axis cs:1237600,-0.5)
--cycle;
\path [overlap]
(axis cs:1271840,-0.5)
--(axis cs:1271840,3.5)
--(axis cs:1287200,3.5)
--(axis cs:1287200,-0.5)
--cycle;
\path [overlap]
(axis cs:1316160,-0.5)
--(axis cs:1316160,3.5)
--(axis cs:1341120,3.5)
--(axis cs:1341120,-0.5)
--cycle;
\path [overlap]
(axis cs:1463200,-0.5)
--(axis cs:1463200,3.5)
--(axis cs:1491040,3.5)
--(axis cs:1491040,-0.5)
--cycle;
\path [overlap]
(axis cs:1557920,-0.5)
--(axis cs:1557920,3.5)
--(axis cs:1574080,3.5)
--(axis cs:1574080,-0.5)
--cycle;

\draw[utt3] (axis cs:30720,-0.4) rectangle (axis cs:204640,0.4);
\draw[utt3] (axis cs:227520,-0.4) rectangle (axis cs:237120,0.4);
\draw[utt3] (axis cs:295360,-0.4) rectangle (axis cs:342240,0.4);
\draw[utt3] (axis cs:348160,-0.4) rectangle (axis cs:480640,0.4);
\draw[utt3] (axis cs:543680,-0.4) rectangle (axis cs:599680,0.4);
\draw[utt3] (axis cs:621440,-0.4) rectangle (axis cs:673120,0.4);
\draw[utt3] (axis cs:676800,-0.4) rectangle (axis cs:758560,0.4);
\draw[utt3] (axis cs:776320,-0.4) rectangle (axis cs:888640,0.4);
\draw[utt3] (axis cs:942880,-0.4) rectangle (axis cs:978560,0.4);
\draw[utt3] (axis cs:982400,-0.4) rectangle (axis cs:998400,0.4);
\draw[utt3] (axis cs:1034080,-0.4) rectangle (axis cs:1044000,0.4);
\draw[utt3] (axis cs:1082240,-0.4) rectangle (axis cs:1171360,0.4);
\draw[utt3] (axis cs:1206240,-0.4) rectangle (axis cs:1237600,0.4);
\draw[utt3] (axis cs:1248480,-0.4) rectangle (axis cs:1287520,0.4);
\draw[utt3] (axis cs:1300480,-0.4) rectangle (axis cs:1309600,0.4);
\draw[utt3] (axis cs:1316160,-0.4) rectangle (axis cs:1341120,0.4);
\draw[utt3] (axis cs:1417920,-0.4) rectangle (axis cs:1450560,0.4);
\draw[utt3] (axis cs:1463200,-0.4) rectangle (axis cs:1492320,0.4);
\draw[utt3] (axis cs:1496800,-0.4) rectangle (axis cs:1574080,0.4);
\draw[utt3] (axis cs:1613600,-0.4) rectangle (axis cs:1655200,0.4);
\draw[utt3] (axis cs:1670240,-0.4) rectangle (axis cs:1697760,0.4);
\draw[utt4] (axis cs:180480,0.6) rectangle (axis cs:205760,1.4);
\draw[utt4] (axis cs:237440,0.6) rectangle (axis cs:288640,1.4);
\draw[utt4] (axis cs:328960,0.6) rectangle (axis cs:346880,1.4);
\draw[utt4] (axis cs:652320,0.6) rectangle (axis cs:670880,1.4);
\draw[utt4] (axis cs:813440,0.6) rectangle (axis cs:828960,1.4);
\draw[utt4] (axis cs:891840,0.6) rectangle (axis cs:898400,1.4);
\draw[utt4] (axis cs:918240,0.6) rectangle (axis cs:946240,1.4);
\draw[utt4] (axis cs:1013760,0.6) rectangle (axis cs:1039520,1.4);
\draw[utt4] (axis cs:1159520,0.6) rectangle (axis cs:1201920,1.4);
\draw[utt4] (axis cs:1271840,0.6) rectangle (axis cs:1287200,1.4);
\draw[utt4] (axis cs:1456320,0.6) rectangle (axis cs:1491040,1.4);
\draw[utt4] (axis cs:1590880,0.6) rectangle (axis cs:1609760,1.4);
\draw[utt2] (axis cs:0,1.6) rectangle (axis cs:21920,2.4);
\draw[utt2] (axis cs:216320,1.6) rectangle (axis cs:227360,2.4);
\draw[utt2] (axis cs:353600,1.6) rectangle (axis cs:363680,2.4);
\draw[utt2] (axis cs:447680,1.6) rectangle (axis cs:456000,2.4);
\draw[utt2] (axis cs:480160,1.6) rectangle (axis cs:548640,2.4);
\draw[utt2] (axis cs:562560,1.6) rectangle (axis cs:573120,2.4);
\draw[utt2] (axis cs:582720,1.6) rectangle (axis cs:622240,2.4);
\draw[utt2] (axis cs:753440,1.6) rectangle (axis cs:763520,2.4);
\draw[utt2] (axis cs:807520,1.6) rectangle (axis cs:838880,2.4);
\draw[utt2] (axis cs:1043680,1.6) rectangle (axis cs:1080320,2.4);
\draw[utt2] (axis cs:1117920,1.6) rectangle (axis cs:1149920,2.4);
\draw[utt2] (axis cs:1155360,1.6) rectangle (axis cs:1170400,2.4);
\draw[utt2] (axis cs:1185280,1.6) rectangle (axis cs:1199680,2.4);
\draw[utt2] (axis cs:1289280,1.6) rectangle (axis cs:1296160,2.4);
\draw[utt2] (axis cs:1330720,1.6) rectangle (axis cs:1338080,2.4);
\draw[utt1] (axis cs:6560,2.6) rectangle (axis cs:63840,3.4);
\draw[utt1] (axis cs:203520,2.6) rectangle (axis cs:221600,3.4);
\draw[utt1] (axis cs:353600,2.6) rectangle (axis cs:363680,3.4);
\draw[utt1] (axis cs:426240,2.6) rectangle (axis cs:437280,3.4);
\draw[utt1] (axis cs:470560,2.6) rectangle (axis cs:501760,3.4);
\draw[utt1] (axis cs:603520,2.6) rectangle (axis cs:624480,3.4);
\draw[utt1] (axis cs:665760,2.6) rectangle (axis cs:691360,3.4);
\draw[utt1] (axis cs:699840,2.6) rectangle (axis cs:710880,3.4);
\draw[utt1] (axis cs:818080,2.6) rectangle (axis cs:838720,3.4);
\draw[utt1] (axis cs:977280,2.6) rectangle (axis cs:1010240,3.4);
\draw[utt1] (axis cs:1231840,2.6) rectangle (axis cs:1237760,3.4);
\draw[utt1] (axis cs:1311840,2.6) rectangle (axis cs:1351840,3.4);
\draw[utt1] (axis cs:1557920,2.6) rectangle (axis cs:1586080,3.4);
\end{axis}

\end{tikzpicture}
    \caption{Activity graph for a meeting segment of the CHIME5 corpus.}
    \label{fig:meeting_example}
\end{figure}

The simulation of speech mixtures that correspond to \cref{eq:signal-model} can be split into three steps:
(i) source data selection (selecting source signals to sample from),
(ii) speaking pattern sampling (sampling sources $\source_\utt$ and their offsets $\offset_\utt$), and
(iii) environment simulation (determining $\scale$, $\rir_\utt$, $\noise$ and possibly additional factors not yet considered in \cref{eq:signal-model}). 
With our mixture signal generator, we provide several building blocks for each of these steps that can be combined flexibly to simulate data of varying scenarios.

The fist step, source data selection, is to select the source databases that provide clean speech signals $\source_\utt$ for mixture creation.
We only require the source databases to contain recordings of clean speech and the speaker identities. All other information, e.g. transcriptions, are passed through unmodified if needed for training purposes later on.
For each mixture, the number of active speakers $\nspk$ and one utterance per active speaker are sampled.

Next, the speaking pattern is sampled.
This sampling differs heavily for classical source separation databases and meetings (see \cref{fig:single_mixture} and \cref{fig:meeting_example}).
In case of classical speech mixtures, only an offset is sampled for each utterance sampled in the source data selection.
MMS-MSG provides sampling functions for the three scenarios depicted in \cref{fig:single_mixture}.
For meeting data, additional utterances per speaker need to be sampled, so that
this stage encompasses additional steps described in \cref{sec:conversational_speaking_pattern}.



Finally, all signal modifications are added to the mixture in the environment simulation.
This stage comprises all postprocessing steps that are necessary to simulate an environment, including source scaling ($\scale_\utt$), reverberation ($\rir_\utt$), additional noise sources ($\noise$), or even additional influences such as sampling rate offsets \cite{22_gburrek_sro}. 
Due to the modular design, the postprocessing steps can be customized and new steps can be added as required.

The simulation process further is divided into a deterministic (random, but fixed through a seed) parameter sampling and the actual mixture creation, so that the parameters can be sampled without loading any actual audio data. 
This enables memory efficient storage, quick prototyping and on-the-fly data generation while ensuring reproducibility.
Moreover, the steps in the simulation process, e.g., sampling the utterances, the offsets or the \glspl{RIR}, are independent, so that each step can be modified while keeping all others constant.
\gls{MMS-MSG} offers native support of dynamic mixing \cite{21_Zheghidour_wavesplit} by varying the seed of the parameter sampling.


    

\subsection{Meeting data simulation}
\label{sec:conversational_speaking_pattern}
To accurately portray the meeting characteristics addressed in \cref{sec:meeting_characteristics} in the simulated meetings while maintaining a high degree of flexibility, two sampling components can be specified for the meeting data generation: speaker-turn sampling and overlap sampling.
We opted for generating a meeting sequentially, i.e., selecting utterances one after another until the specified meeting length is reached.

\subsubsection{Speaker-turn sampling}
The speaker-turn sampling determines the speaking order in a meeting (i.e., which speaker's turn it is next).
In its most basic form, the next active speaker can be chosen at random or in a round robin fashion.
However, these basic approaches do not give control over the distribution of the activity per speaker. 

We propose an activity-based speaker turn sampling. 
Here, the probability $\prob_\utt(\spk)$ that speaker $\spk$ is active for the next utterance $\utt$ is the normalized inverse activity of this speaker:
\begin{align}
	\prob_\utt(\spk) = \frac{\frac{1}{\estactivity_{\spk\utt}}}{\sum_{\spk=1}^\nspk\frac{1}{\estactivity_{\spk\utt}}}.
\end{align}
The activity $\estactivity_{\spk\utt}$ is the share of activity of speaker $\spk$ up to this point in the meeting. 
As each speaker is weighted by its inverse activity,
speakers are less likely to be active if they already have a high activity in the meeting, resulting in a roughly equal activity of each speaker over the course of the meeting.
By replacing the current activity $\estactivity_\spk$ with the difference to the desired activity $\activity_\spk$ and clipping at \num{0}, any other activity distribution, such as highly asymmetrical conversations (e.g., a lecture scenario), also can be simulated.
On a side note, this speaker-turn sampling encourages the occurrence of monologues if a single speaker is underrepresented activity-wise. As it is not unlikely for a speaker to be active for multiple consecutive utterances in real meetings, this side effect further helps to increase the realism of the simulated meetings.

After the speaker is selected, an utterance has to be chosen for that speaker. 
This is done by first grouping the utterances from the source database by the speaker and optionally by a scenario, e.g., to keep the acoustic conditions for a speaker similar within a meeting.
Then, one group is selected for each speaker in the meeting and utterances are sampled randomly, making sure that all utterances appear once before re-using an utterance.
    
\begin{figure}[bt]
    \centering
    \input{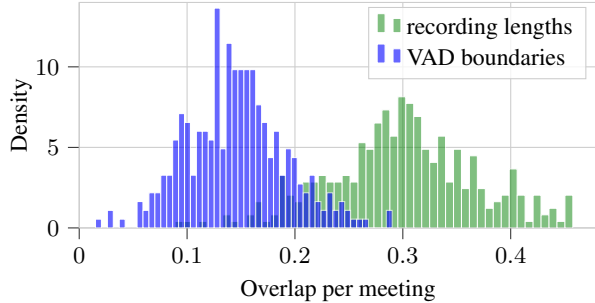}
    \caption{Probability density of the overlap per WSJ meeting for the \enquote{medium overlap} scenario for a calculation based on the utterance or VAD boundaries.}
    \label{fig:overlap_boundaries}
\end{figure}
    
\subsubsection{Overlap Sampling}
After deciding on the speaker and selecting an utterance $\source_\utt$, an offset $\offset_\utt$ is sampled for this utterance.
It is first randomly selected whether silence or overlap occurs.
In case of silence, the duration is sampled and the next speaker-turn sampling proceeds.
In case of overlap, first the maximally allowed overlap is determined before randomly sampling an overlap length. 
This maximally allowed overlap is determined by the given maximum while ensuring the specified number of concurrent speakers (i.e. the number of simultaneously active speakers) is not exceeded.
Both the silence and overlap ranges are denoted as the absolute duration (in seconds or samples) between two consecutive utterances, not as the desired total overlap per meeting.
This is due to the fact that the single-speaker utterances from most source datasets typically begin and end with some silence, so that the desired overlap cannot be accurately sampled based on the recording lengths.
\Cref{fig:overlap_boundaries} illustrates this problem by plotting the distribution of overlap for a meeting determined with the recording lengths compared to the actual overlap computed with a \gls{VAD}.
The real overlap is significantly smaller. 
However, this problem can be compensated for by setting the overlap lengths in absolute values and introducing a minimal overlap. 
Here, a source data specific minimal overlap can be set to account for the silence regions at the borders of each utterance.



\begin{figure}[bt]
    \centering
\begin{tikzpicture}

    \definecolor{darkslategray38}{RGB}{38,38,38}
    \definecolor{gray}{RGB}{128,128,128}
    \definecolor{green}{RGB}{0,128,0}
    \definecolor{lightgray204}{RGB}{204,204,204}
    \definecolor{orange}{RGB}{255,165,0}
    
    \begin{axis}[
        scale only axis,
        width=0.85\columnwidth,
        height=2cm,
        axis line style={rounded corners=3,draw=none},
        axis background/.style={uttbox},
        tick align=outside,
        xtick pos=bottom,
        xmin=0, xmax=950000,
        xtick={0,240000, 480000, 720000, 960000},
        xticklabels={0,30,60,90,120},
        xminorgrids,
        xtick style={color=darkslategray38},
        y grid style={lightgray204},
        ymin=-0.5, ymax=3.5,
        ytick style={color=darkslategray38},
        scaled x ticks=base 10:0,
        xlabel={s},
        xlabel style={
            at={(ticklabel cs:1)},
            anchor={south east}
        },
        ytick={0,1,2,3,4},
        yticklabels={D, C, B, A},
        ylabel={Speaker},
        ylabel shift=-4pt,
        xlabel={time / s},
        ytick pos=left,
    ]
    \path [overlap]
    (axis cs:133977,-0.5)
    --(axis cs:133977,3.5)
    --(axis cs:138738,3.5)
    --(axis cs:138738,-0.5)
    --cycle;
    \path [overlap]
    (axis cs:221077,-0.5)
    --(axis cs:221077,3.5)
    --(axis cs:238485,3.5)
    --(axis cs:238485,-0.5)
    --cycle;
    \path [overlap]
    (axis cs:267811,-0.5)
    --(axis cs:267811,3.5)
    --(axis cs:319320,3.5)
    --(axis cs:319320,-0.5)
    --cycle;
    \path [overlap]
    (axis cs:340422,-0.5)
    --(axis cs:340422,3.5)
    --(axis cs:357510,3.5)
    --(axis cs:357510,-0.5)
    --cycle;
    \path [overlap]
    (axis cs:389137,-0.5)
    --(axis cs:389137,3.5)
    --(axis cs:413025,3.5)
    --(axis cs:413025,-0.5)
    --cycle;
    \path [overlap]
    (axis cs:452063,-0.5)
    --(axis cs:452063,3.5)
    --(axis cs:453487,3.5)
    --(axis cs:453487,-0.5)
    --cycle;
    \path [overlap]
    (axis cs:464628,-0.5)
    --(axis cs:464628,3.5)
    --(axis cs:486801,3.5)
    --(axis cs:486801,-0.5)
    --cycle;
    \path [overlap]
    (axis cs:538010,-0.5)
    --(axis cs:538010,3.5)
    --(axis cs:559091,3.5)
    --(axis cs:559091,-0.5)
    --cycle;
    \path [overlap]
    (axis cs:747889,-0.5)
    --(axis cs:747889,3.5)
    --(axis cs:751448,3.5)
    --(axis cs:751448,-0.5)
    --cycle;
    \path [overlap]
    (axis cs:802922,-0.5)
    --(axis cs:802922,3.5)
    --(axis cs:821534,3.5)
    --(axis cs:821534,-0.5)
    --cycle;
    \path [overlap]
    (axis cs:846467,-0.5)
    --(axis cs:846467,3.5)
    --(axis cs:856882,3.5)
    --(axis cs:856882,-0.5)
    --cycle;
    \path [overlap]
    (axis cs:877976,-0.5)
    --(axis cs:877976,3.5)
    --(axis cs:922031,3.5)
    --(axis cs:922031,-0.5)
    --cycle;
    \draw[utt1] (axis cs:0,2.6) rectangle (axis cs:43103,3.4);
    \draw[utt1] (axis cs:340422,2.6) rectangle (axis cs:413025,3.4);
    \draw[utt1] (axis cs:452063,2.6) rectangle (axis cs:486801,3.4);
    \draw[utt1] (axis cs:538010,2.6) rectangle (axis cs:597969,3.4);
    \draw[utt1] (axis cs:607385,2.6) rectangle (axis cs:637307,3.4);
    \draw[utt1] (axis cs:639788,2.6) rectangle (axis cs:713888,3.4);
    \draw[utt1] (axis cs:726110,2.6) rectangle (axis cs:751448,3.4);
    \draw[utt1] (axis cs:802922,2.6) rectangle (axis cs:856882,3.4);
    \draw[utt4] (axis cs:54714,0.6) rectangle (axis cs:138738,1.4);
    \draw[utt4] (axis cs:389137,0.6) rectangle (axis cs:453487,1.4);
    \draw[utt4] (axis cs:747889,0.6) rectangle (axis cs:821534,1.4);
    \draw[utt4] (axis cs:877976,0.6) rectangle (axis cs:963525,1.4);
    \draw[utt2] (axis cs:133977,1.6) rectangle (axis cs:238485,2.4);
    \draw[utt2] (axis cs:266776,1.6) rectangle (axis cs:357510,2.4);
    \draw[utt3] (axis cs:221077,-0.4) rectangle (axis cs:255189,0.4);
    \draw[utt3] (axis cs:267811,-0.4) rectangle (axis cs:319320,0.4);
    \draw[utt3] (axis cs:464628,-0.4) rectangle (axis cs:559091,0.4);
    \draw[utt3] (axis cs:846467,-0.4) rectangle (axis cs:922031,0.4);
    \end{axis}
    
    \end{tikzpicture}
    
    \caption{Speaker activity for one example meeting with the \enquote{medium overlap} configuration.}
    \label{fig:example_meeting}
\end{figure}
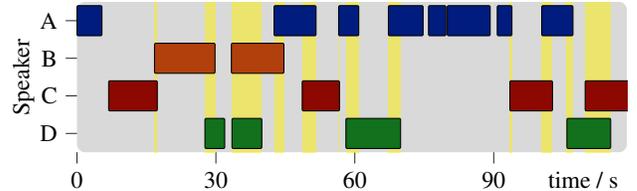

\section{Experiments}
\label{sec:experiments}
For evaluation purposes, we provide a test meeting dataset based on WSJ \cite{Garofalo2007_CsriWsj0Complete} at a sample rate of \SI{8}{\kilo\hertz}. We provide three different overlap configurations which are inspired by the scenarios of LibriCSS, one for no overlap, medium overlap, and high overlap, each. 
The parameters used for sampling and the resulting amount of overlap and silence are depicted in \cref{tab:test_conditions}.
Additionally, in the high overlap scenario, the probability of silence is decreased from \SI{10}{\percent} to \SI{1}{\percent} to ensure a higher amount of overlap.
An example activity for the normal overlap scenario is plotted in \cref{fig:example_meeting}.
Compared to the Chime-5 database, the source utterances in WSJ are much longer so that fewer speaker turns are present in the same time length.
In each scenario, \num{16} meetings of roughly 2 minutes are sampled for \num{5}-\num{8} speakers, each, resulting in \num{64} meetings (i.e. \SI{2}{\hour} of audio) per scenario. 
All three scenarios use the same configuration except for the overlap sampling so that the scenarios only differ in the amount of overlap and number of utterances, but not in utterance order or environment simulation.
In this way, it is possible to accurately determine the impact an increasing overlap has on the speech recognition performance. 
For the reverberation, room impulses simulated with the image method proposed in \cite{allen1979image} were used. 
The room configurations of the simulated rooms and parameters for source scaling and noise were taken from \cite{Drude_19_smswsj}, albeit with \num{8} instead of \num{4} speaker positions per room.

\begin{table}[bt]
    \centering
        \caption{Meeting conditions for the WSJ Meeting test scenarios. ov and sil are the parameters for overlap sampling. $\mathrm{ov}_{\mathrm{rel}}$ and $\mathrm{sil}_{\mathrm{rel}}$ denote the measured overlap or silence based on VAD boundaries.}
    \label{tab:test_conditions}
    \begin{tabular}{l c c c c}
    \toprule
                 &    \multicolumn{2}{c}{Parameters} & \multicolumn{2}{c}{Measured} \\
                 \cmidrule(lr){2-3}\cmidrule(lr){4-5}
      Scenario   &  $\mathrm{ov}$ [\si{\second}] & $\mathrm{sil}$ [\si{\second}] & $\mathrm{ov}_{\mathrm{rel}}$ [\si{\percent}] & $\mathrm{sil}_{\mathrm{rel}}$ [\si{\percent}]\\
    \midrule
    no ov & 0-0 & 0-2 & 0.00 & 32.8 $\pm$ 23.0 \\
    medium ov & 0-8 & 0-2  & 15.8 $\pm$ 4.3 & 18.7 $\pm$ 3.3\\
    high ov & 2-8 & 0-1  & 25.9 $\pm$ 4.3  & 14.7 $\pm$ 3.3\\
    \bottomrule
    \end{tabular}
\end{table}

\subsection{Baseline Recipe}
\label{sec:baseline}

The baseline model is a single-channel source separation network which uses the \gls{CSS} pipeline \cite{20_Chen_libricss}. 
The pipeline segments the recording into overlapping chunks of \SI{4}{\second}, processes each chunk separately with the source separator and then stitches them back together to obtain two output streams for the recording. 
For speech recognition, the separated audio streams are partitioned into single utterances by a \gls{VAD} module before they are passed to an \gls{ASR} system. 

The source separation network is a simple PIT-BLSTM \cite{Kolbaek_17_uPit} consisting of \num{3} \gls{BLSTM} layers with \num{600} bidirectional units followed by \num{2} fully connected layers. As training loss, the \gls{SASDR} \cite{21_Neumann_sa_sdr} with a threshold of \SI{25}{\decibel} is used.
For stitching, the MSE-based approach of \cite{20_Chen_libricss} is used with a history context, content window and future context of \SI{1.2}{\second}, \SI{2.4}{\second} and \SI{0.4}{\second}, respectively.
For \gls{ASR}, the model from \cite{Drude_19_smswsj} is used. It consists of a  \gls{TDNN-F} that is trained on reverberated WSJ utterances. 

For evaluation, we compute the utterance-level \gls{SDR} \cite{06_Vincent_sdr} averaged over all scenarios and the \gls{ORC-WER} \cite{Sklyar2021_MultiturnRNNTStreaming} for each scenario. 
The utterance-level \gls{SDR} measures the signal-level source separation quality. In the context of meeting-data, this metric also indicates how many distortions are introduced to single-speaker regions.
The \gls{ORC-WER} is a diarization-agnostic way to measure the \gls{WER} of the system. 
It is the minimum \gls{WER} among the possible combinations of reference transcriptions matched with the estimated output stream transcriptions.

\subsection{Results}

\cref{tab:wsj_anechoic} and \cref{tab:wsj_reverb} show the \gls{WER} for our baseline model and the averaged \gls{SDR} over all scenarios for anechoic and reverberant test data, respectively.
The model either is trained on full two-speaker mixtures of the SMS-WSJ scenario, chunks of these mixtures or chunks of a simulated meeting. The chunk size was set to \SI{4}{\second} to match the stitching parameters.
Here, it becomes apparent that, especially for reverberant data, the model trained on full overlap mixtures introduces many distortions in single-speaker regions, resulting in a high \gls{WER} for the scenario without overlap.
Using mixture segments for training helps to prevent these distortions, as  single-speaker signals also are processed during training of these models. 
While the model trained on meeting segments, i.e. in a matched condition, shows a higher \gls{SDR} for all scenarios, the \gls{WER} only improves for test conditions of lower overlap. This indicates that during training, the overlap must be chosen higher for the model to perform a good separation. Also, a training in reverberant conditions profits more from a high overlap than one for anechoic data. 

While the differences are not large, they show that the usage of the training scenario impacts the environments the model can be deployed for. 
This again highlights one main benefit of MMS-MSG. Using the mixture generator, multiple scenarios can be simulated that only differ in a single simulation component. Therefore, it is easy to directly investigate the impact of environmental influences (e.g. reverberation) or varying degrees of overlap.  Also, the possibility of providing conversational training data allows the usage of training losses like Graph-Pit \cite{vonNeumann2021_GraphPITGeneralizedPermutation} that require access to this data.

\begin{table}
    \sisetup{detect-weight}
	\robustify\bfseries  
    \sisetup{round-precision=2,round-mode=places, table-format = 2.1}
    \centering
        \caption{Results for the anechoic WSJ meeting dataset for different training data configurations.}
    \label{tab:wsj_anechoic}
    \begin{tabular}{lSSSS}
         \toprule
         Training Data    &{$\overline{\mathrm{SDR}}$} & \multicolumn{3}{c}{ORC WER}\\
             & {[\si{\decibel}]}& {no ov} &  {normal ov} & {high ov} \\
         \midrule
         No separation & 11.5116 & 7.3664941295291 & 33.25245930565971 & 43.836356619065997 \\
         Full overlap & 18.664 & 8.161848251483399 &  16.564471304424544 & 21.096101891161714\\ %
         Full overlap (chunk) & 18.796 & 7.562176492867062 &  15.673386563147912 & 20.12736395214203\\ 
         Meetings (normal ov) & 19.925 & 7.045880149812735 & 15.488111517931977 & 19.891933616364338 \\ 
         Meetings (high ov) & 19.57 & 6.798384042418887 & 15.188111517931977 & 20.463 \\ 
    \bottomrule
    \end{tabular}

\end{table}

\begin{table}
    \sisetup{detect-weight}
	\robustify\bfseries  
    \sisetup{round-precision=2,round-mode=places, table-format = 2.1}
    \centering
        \caption{Results for the reverberant WSJ meeting dataset for different training data configurations.}
    \label{tab:wsj_reverb}
    \begin{tabular}{lSSSS}
         \toprule
         Training Data   &{$\overline{\mathrm{SDR}}$} & \multicolumn{3}{c}{ORC WER}\\
          & {[\si{\decibel}]}  & {no ov} &  {normal ov} & {high ov} \\
         \midrule
         No separation & 6.879 & 9.771493498295669 & 36.296263619921476 & 46.78502508683906\\ 
         Full overlap &12.961 & 18.267895467743972 &  24.672460187921832  & 28.776534156696254 \\ 
         Full overlap (chunk) & 13.207 & 10.453  & 23.230 & 30.012\\ 
        Meeting (normal ov)  & 13.421  & 9.803055169801793 &  23.313776523004984 & 30.405248938633733\\ 
        Meetings (high ov) & 13.55467 & 9.329630097209948 & 22.574043848427367  & 30.215389273 \\ 
    \bottomrule
    \end{tabular}

\end{table}

\subsection{Evaluation on LibriCSS}
To further demonstrate the ability of \gls{MMS-MSG} to be used for the system design, in particular for realistic scenarios, we evaluate our baseline model on the LibriCSS database.
Compared to the training data from \cref{sec:baseline}, only the source dataset is changed from WSJ to LibriSpeech and the room impulse responses are generated for a sampling rate of \SI{16}{\kilo\hertz}. Again, the PIT-BLSTM separator is trained on the simulated meetings and evaluate with the \gls{CSS} pipeline.
As the \gls{ASR} system, the pretrained the ESPnet \cite{Watanabe2018_ESPnetEndtoEndSpeech} system from \cite{watanabe2020PretrainedASR} is used as it is easily applicable and more robust to artefacts than the LibriCSS baseline model, while providing a similar performance on single-speaker regions.
\cref{tab:libricss} shows the results for our baseline system that was trained with the \gls{MMS-MSG}  training data simulation. The results outperform those obtained with the model used in \cite{20_Chen_libricss}. While the results can not be directly compared, they nevertheless show that the meetings simulated by \gls{MMS-MSG} are at least equally suited for training as the LibriSpeech mixtures from \cite{20_Chen_libricss}. 

\begin{table}[bt]
    \sisetup{detect-weight}
	\robustify\bfseries  
    \sisetup{round-precision=1,round-mode=places, table-format = 2.1}
    \setlength\tabcolsep{5pt}
    \centering
        \caption{Result of our baseline system on the LibriCSS datasets 0S\=/OV40 for simulated LibriSpeech meetings as training data}
    \label{tab:libricss}
    \begin{tabular}{lSSSSSS|S}
    \toprule
    Model & \multicolumn{7}{c}{{WER}} \\
    & {0S} & {0L} & {OV10} & {OV20} & {OV30} & {OV40} & {avg.} \\
    \midrule
    \cite{20_Chen_libricss} & 17.6 & 16.3 & 20.9 & 26.1 & 32.6 & 36.1 & 24.933 \\
    Ours & 11.891988555078683 & 19.549365014338385  & 13.191179762704275 & 18.45440988495952 & 23.881327719573093 & 26.973560847552974 & 18.9895\\
    \bottomrule
    \end{tabular}

\end{table}
\section{Conclusions}
\label{sec:summary}
We provide an open-source simulation tool for a multitude of scenarios. The data is simulated on-demand and provides access to all intermediate signals, so that all single system components can be evaluated. While our experiments focus on source separation, our framework also can be used for the training or evaluation of diarization or speech enhancement components.
 Thus, our framework allows the finetuning and evaluation of every single component of a transcription system on simulated meetings  before switching over to real recordings like CHIME-5/6, enabling a more fine-grained analysis of system components.
While we also provide a meeting-scenario test set based on  WSJ data, the focal point of \gls{MMS-MSG} lies in its versatility.
Using a highly modular structure, it is the successor of SMS-WSJ and aims to provide a uniform framework for meeting data generation of all necessary scenarios.
Therefore, the exemplary investigation on the impact of the overlap on the speech recognition quality can be easily transferred to any other source database, e.g. LibriSpeech. Furthermore, the design of new test sets, e.g. with asymmetric activity distributions, is possible and just as easy.
In the future, we will integrate more components like HMM-based speaking-order sampling as in  \cite{22_yamashita_dia_data} 
to allow for an even higher versatility and realism.



\section{Acknowledgements}
Computational resources were provided by the Paderborn Center for Parallel Computing.
Funded by the German Research Foundation (DFG, Deutsche Forschungsgemeinschaft) - Project no. 282835863.
\vfill\pagebreak

\bibliographystyle{IEEEbib}
\bibliography{refs}

\end{document}